# Controlling the long-term dynamics of Brillouin Fiber Lasers using pump modulation: Bi-stable lasing and Hysteresis

OMER KOTLICKI AND JACOB SCHEUER[*]

*School of Electrical Engineering, Tel-Aviv University, Tel-Aviv 69978, Israel*
*Corresponding author: kobys@eng.tau.ac.il*

**The long-term lasing dynamics of Brillouin Fiber Lasers was recently shown to be governed by an intrinsic thermal feedback which stabilizes the lasing frequency at the lower half of the gain line. We show that this feedback can be utilized by means of pump modulation to greatly control this dynamics. Specifically, we demonstrate experimentally a bi-stable fiber laser where switching between lasing modes is possible by either controlling the cavity length or pump power. An inherent hysteretic response ensures the long-term stability of the laser in each mode. Excellent agreement is found between the experimental results and a master-equation based theoretical model.**

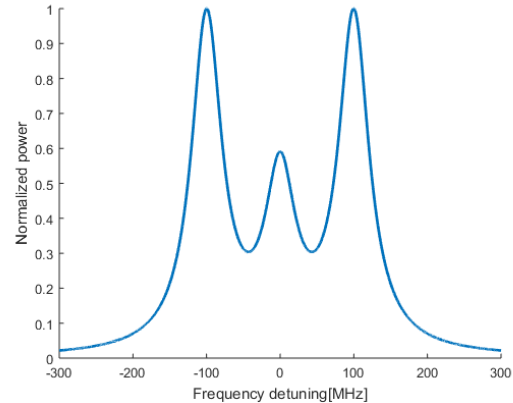

Fig 1. (color online) A simulated line-shape of SBS with pump modulated with a frequency of 100MHz and depth of 80%. The frequency detuning is relative to the peak of the Brillouin gain-line generated by an unmodulated pump signal.

Brillouin Fiber Lasers (BFLs) utilize Stimulated Brillouin Scattering (SBS) in optical fibers for gain [1–3]. Such lasers are known for their very low thresholds and ultra-narrow linewidths but are plagued by a host of internal and environmental parameters which affect their long-term stability [4]. Two important factors contribute to the lasing dynamics of BFLs [5]: The first factor is losses inside fiber cavities which convert part of the optical power to heat. The heat alters the refractive index of the cavity material (through the thermo-optic effect and thermal expansion) and hence affects the optical path of light travelling inside the cavity [6,7]. The second factor, common to many laser types, is the dependence of the lasing power on the frequency detuning between the lasing frequency and the peak of the gain-line. Maximum lasing power is obtained when the two overlap and as the detuning grows, the lasing power drops accordingly [8]. The interaction between these two factors constitutes an inherent thermal feedback linking between the lasing power and frequency of BFLs. As previously reported [5], this feedback actively stabilizes the lasing power and frequency of BFLs, when lasing occurs on the lower frequency half of the gain-line; however, on the higher frequency half of the gain-line, this feedback destabilizes the lasing process constantly pushing the lasing frequency to the stable lower frequency half of the gain-line.

The properties of this feedback mechanism depend directly on the spectral shape of the gain-line. Thus, reshaping the gain-line can provide a powerful tool for controlling the feedback mechanism and, hence, the BFL dynamics. Pump modulation had been shown to affect the intrinsic line-shape of SBS and was used to demonstrate slow light related effects [9,10]. In additions, pump modulation has been used to improve the stability of BFLs [11]. Sinusoidal modulation of signals result in frequency sidebands (AM modulation) [12]. Such modulated signals can be used to pump BFLs resulting in a combined gain-line shape of three intrinsic Lorentzian SBS lines where the inner one originates from the original pump and the outer two are the modulation sidebands. The modulation frequency determines the frequency difference between the peaks of neighboring Lorentzians and the modulation depth determines the ratio between the strength of the internal and external Lorentzians. Fig. 1 shows an example of a spectral profile of the Brillouin gain due to a modulated pump where the modulation depth and frequency are 80% and 100MHz respectively. The SBS linewidth of each Lorentzian is taken as 50MHz, which is a typical value for silica fibers.

In this paper, we use a modulated pump to enable lasing at two separate frequency bands and utilize the thermal feedback to achieve bi-stable lasing. In each frequency band, the lasing parameters are kept stable by the feedback mechanism (dashed green regions in fig 2.a.) while switching between the two bands is made possible using external perturbations of either the optical length of the cavity or the pump power.

Fig 2.a shows the spectral regions of the Brillouin gain (solid blue) where the thermal feedback stabilizes the lasing process (dashed green) along with the regions where it destabilizes the process (dotted red). Fig. 2.b shows an overlay of the cold cavity modes on top of the gain line. The difference between the Free Spectral Range (FSR) of the cavity and the modulation frequency ensures that only a single mode can saturate the gain and lase.

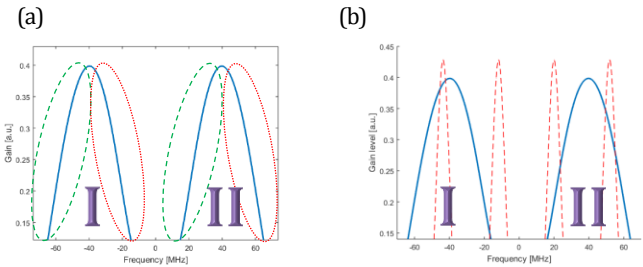

Fig. 2. (color online) An overview of the lasing process. The solid blue line indicates two Lorentzian SBS gain-line peaks spaced 80MHz apart, marked in Roman numerals. (a) shows the regions where stable (dashed green region) lasing and unstable (dotted red region) exist (b) shows a schematic overlay of the cold cavity modes (dashed red curve) over the gain profile.

Lasing occurs at the cavity mode which experiences the largest gain. As the frequency comb of the cold cavity modes shifts in response to various perturbations, mode hopping may occur. Given the spacing between the two Lorentzian gain lines relative to the FSR of the cavity, mode hopping can only take place between two non-adjacent modes on each Lorentzian gain line. Fig. 3 shows the depleted pump level following a single round trip inside the cavity under a fast linear perturbation to the optical length of the cavity (which overcomes the thermal feedback). This measurement is inversely proportional to the shape of the gain line [5] where the two concave dips emanate from the two SBS gain line peaks (see fig. 2) and the mode hopping point between the two gain lines is seen in the non-continuous point at the center of the figure. When the system is perturbed at a slow rate, the thermal feedback discussed in [5] can compensate the perturbation and generate spectral bands where stable lasing takes place (marked by dashed green lines in Fig. 3) as well as regions of instability (marked by dotted red lines in Fig. 3).

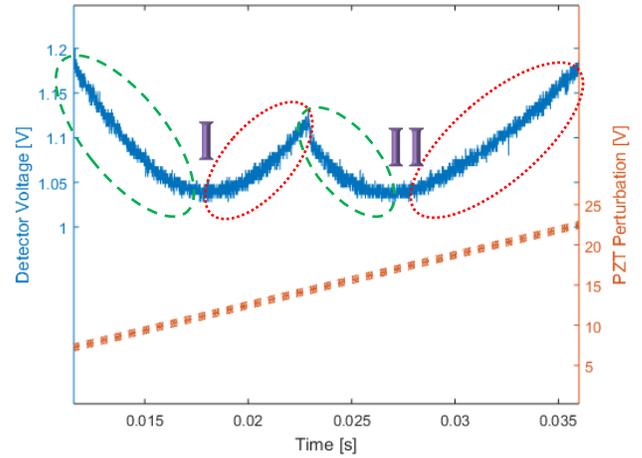

Fig 3. (color online) The depleted pump level following a single round trip inside the cavity as a function of time, given a fast linear perturbation which alters the optical length of the cavity by a single wavelength (shown in the thick dotted orange line). The Roman numerals mark the specific Lorentzian in correspondence with fig. 2. The regions where stable lasing is possible are marked using dashed green lines and the regions where lasing is unstable are marked using dotted red lines.

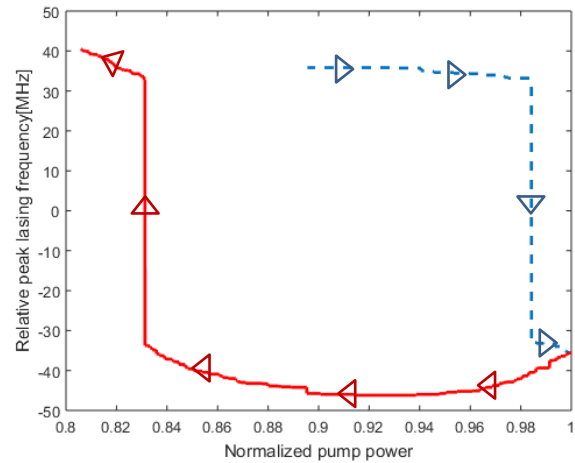

Fig. 4. (color online) Calculated peak lasing frequency in response to a perturbation consisting of a positive ramp in pump power (dashed blue curve) followed by a negative ramp in pump power (solid red curve). See also Visualization 1.

A master-equation based model was used to model the dynamic properties of the system. A comprehensive description of this model had previously been given [5]. The main difference between the previously published model and the one used here is the spectrum of a gain line which consists of two Lorentzian lines spaced 80MHz apart. Fig. 4 depicts the dynamic response of the lasing frequency to perturbations in the pump power. The initial lasing frequency is in the higher frequencies stability band (Lorentzian II in Fig. 2 and 3) when pump power is increased linearly from 90% of its maximal value to 100% (dashed blue curve). As shown in Fig 4, the lasing frequency hops to the lower frequencies stability region (Lorentzian I in Fig. 2 and 3) when the pump power reaches ~98%. After the pump power reaches 100%, the pump is decreased linearly to ~80% (solid red curve). Initially, the thermal feedback process still reacts to the preceding increase in pump power due to the relatively long time constant of the mechanism [5]. Consequently, the lasing frequency continues to decrease, albeit at a lower rate. After a short while the trend is reversed, the lasing frequency starts shifting upward as a direct result of the decrease in pump power and when pump power reaches ~83%, the lasing frequency hops back to the higher frequency stability region (Lorentzian II). Note that the hop from the higher frequency band to the lower one occurred at ~98% pump power while the hop from the lower band to the higher one occurred at ~83%. This signifies a hysteresis which results from the combination of the gain profile and the thermal feedback.

Fig. 5 presents a schematic of the experimental BFL setup. A stable, narrow-line 1550nm DFB laser source is modulated using an Electro Optic Modulator (EOM) and then amplified and employed as a pump source for the BFL. The pump is injected into a short single mode fiber ring resonator (~6m long) using an optical circulator which prevents the pump from resonating in the fiber cavity as it absorbs counter-clockwise travelling waves. On the other hand, the resulting back scattered Brillouin signal can resonate in the cavity as clockwise travelling waves can pass through the circulator. A piezoelectric fiber stretcher is used for introducing variations in the optical length of the cavity and a 1:99 fiber coupler taps the Brillion lasing signal. A 1:1 coupler beats the tapped Brillouin signal with the pump and the combined signal is inserted into a heterodyne detection scheme consisting of a fast optical detector connected to a high resolution RF spectrum analyzer. In the counter-clockwise direction, the 1% tap also samples the depleted pump before it is absorbed by the circulator. The depleted pump level is used for the measurement depicted in Fig. 3 and for the calibration process of the system. The resonator is covered using a polystyrene box which provides sufficient isolation from environmental perturbations.

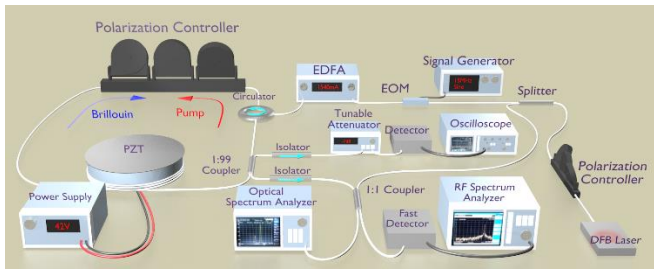

Fig. 5. (color online) A schematic of the experimental setup. The PZT - Piezoelectric Transducer, EDFA - Erbium Doped Fiber Amplifier, EOM - Electro-Optic Modulator.

Perturbations to the BFL pump power are generated by altering the EDFA pump current. Fig. 6 shows the results of two experiments where the pump power was swept upward and then downward. The peak of the lasing line is measured directly using the RF spectrum analyzer and a hysteresis of 140-150mA is clearly visible. The prominent difference between the results of the two experiments is in the region where the pump sweep changes direction (the meeting point of the dashed blue curve and the solid red curve) and emanates from a different starting position relative to the peak of the relevant Lorentzian gain line: in fig 6.a, the starting position is relatively far from the peak of the (higher frequency) Lorentzian gain-line and hence the hop to the lower frequency region occurs at a lower pump current relative to fig 6.b. This is also manifested by the different dynamics of the lasing frequency, stemming from the interaction between the sweep direction and the thermal feedback process.

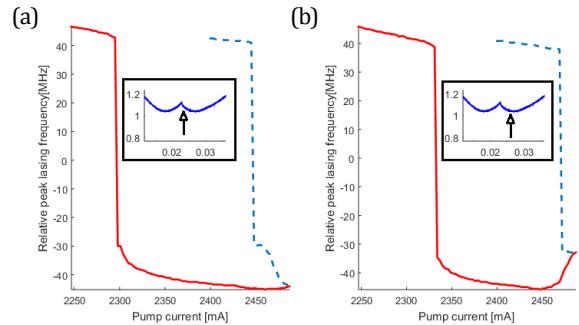

Fig. 6. (color online) Two experimental measurements of the peak lasing frequency relative to the pump power in response to a perturbation consisting of a positive ramp in pump power (dashed blue curve) followed by a negative ramp in pump power (solid red curve). The arrows at the insets mark the different starting lasing frequencies, relative to the gain-line portrayal of fig 3.

Visualization 1 contains a video demonstration of the bi-stability of the system and the behavior of the thermal feedback in response to small step perturbations in the optical length of the cavity. The video shows a synchronized overlay of the RF spectrum analyzer showing the heterodyne down converted laser spectrum and a scope output displaying the applied PZT voltage. Throughout the demonstration, the system responds to manually applied step perturbations. During the first 6 seconds, lasing takes place in the stable region of Lorentzian I. The lasing frequency shifts in response to the applied perturbations and immediately afterwards, the thermal feedback process counteracts the perturbations by shifting the lasing frequency towards its previous location. In the 7th second, a stronger perturbation switches the lasing frequency to the stable region of Lorentzian II. Seconds 8-16 show perturbations similar to the first 6 seconds but at Lorentzian II, demonstrating that this is a stable lasing point. In seconds 17-26, lasing is switched back to Lorentzian I and similar dynamics is shown once again. We note that the heterodyne beating process between the BFL output and the higher frequency pump laser results in a mirror image of the actual laser spectrum and hence Lorentzian I appears on the right side of the displayed spectrum and vice versa.

It is important to note that the BFL cavity comprises several different fiber components which exhibit minor differences in characteristics. As a result, each fiber has a slightly different Brillouin shift and interaction strength. Splicing these fiber sections together yields a complex intrinsic SBS gain profile which consists of several shifted Lorentzians, each with a different peak frequency and different magnitude (which also depends on the length of each specific fiber). As mentioned above, the modulation of the pump results in three copies of the intrinsic SBS line (see fig. 1). As the intrinsic SBS line-shape is complex, the modulation frequency and depth control a superposition of three copies of the complex SBS profile. The modulation parameters were set to achieve a combined SBS gain profile with two dominant peaks of similar magnitude and shape, as required for this work and demonstrated on fig 3.

As shown, pump modulation can alter the line-shape of SBS. Since the dynamics of the thermal feedback depend on the slope of the gain curve, it is possible to use pump modulation to alter the feedback's properties and even substantially reduce it. One place where feedback reduction can be beneficial is sensing applications, where accurate and repeatable responses to perturbations are imperative and by flattening the gain-line, one can greatly reduce the thermal feedback while maintaining single mode lasing across a single FSR around the gain-line center [13]. Fig. 7 shows the depleted pump level following a single round trip inside the cavity (similar to fig. 3). The blue curve is proportional to the gain line of an *unmodulated* BFL where the steep slopes indicate strong feedback (note the slight asymmetry emanating from the presence of several spliced fibers). The orange curve is proportional to the gain line of a BFL where pump modulation was set to produce a flatter gain line while maintaining single mode lasing and the gentle slopes indicate the presence of a substantially weaker thermal feedback.

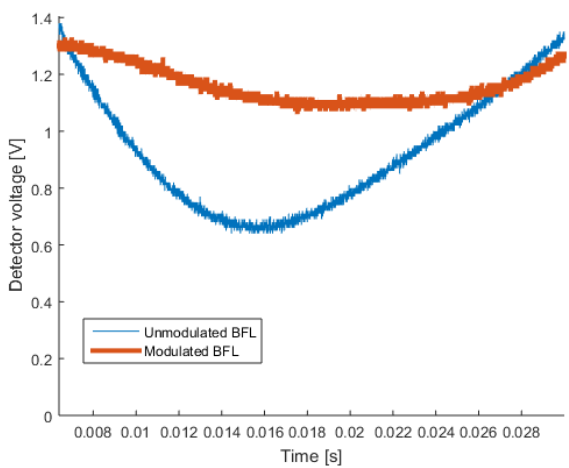

Fig. 7. (color online) The depleted pump level following a single round trip inside the cavity for an unmodulated pump (thin blue curve) and for pump modulation set with an aim to flatten the gain line while maintaining single mode lasing (thick orange curve).

As stronger feedback results in a quicker response, the strength of the thermal feedback can be assessed by measuring the maximal change in output power to small-signal saw-tooth perturbations in the optical length of the cavity. Fig. 8 shows the results of such measurements using unmodulated (thick blue circles) and modulated (thin red circles) pumps. As previously reported [5], the system behaves as a high pass filter and by fitting the results to a typical response of a single pole high pass filter $[(\nu/\nu_0)/(1+\nu/\nu_0)]$, it is possible to extract the characteristic time constant of the feedback system. The feedback pole frequency in the unmodulated case is measured as $\nu_{0,unmodulated}$=*0.721Hz* while in the modulated case it is $\nu_{0,modulated}$=*0.336Hz*, corresponding to time constants of 1.38s and 2.98s respectively, yielding a reduction of 56% in the strength of the feedback.

In conclusion, we studied the impact of pump modulation on the properties of the recently discovered thermal feedback mechanism. We showed the ability to obtain bi stable lasing in BFLs where switching between the two stable frequency regions can be performed using either changes to the pump power or the optical length of the cavity. The approach was verified experimentally and simulated using a dynamic, master-equation based model with excellent agreement between the two. Lastly, we show that pump modulation can also be used to greatly reduce the effect of the thermal feedback by more than a factor of two. This method could be useful for sensing applications requiring consistent and accurate responses to perturbations.

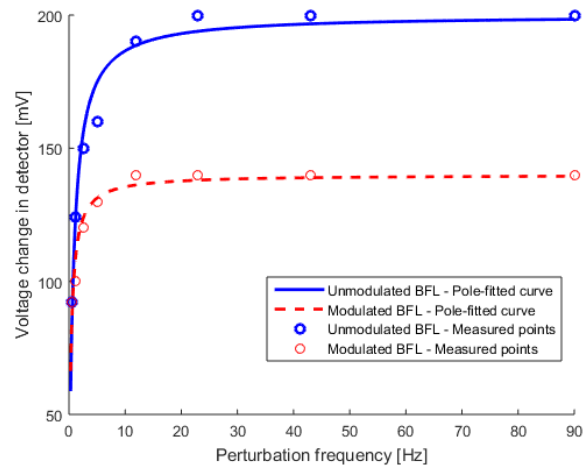

Fig. 8. (color online) The response of the unmodulated (solid blue) and modulated (dashed red) BFLs to small-signal saw-tooth perturbations of different frequencies. The circles indicate the experimental results while the lines portray a fitted high pass filter curve.